\documentclass[reprint,floatfix,notitlepage,nofootinbib,twocolumn,superscriptaddress,pra]{revtex4-1}
\usepackage{amsmath}
\usepackage{amssymb}
\usepackage{graphicx}
\usepackage[tight]{subfigure}
\usepackage{enumitem}
\usepackage{soul}
\usepackage{pifont}
\usepackage{lipsum}
\usepackage{xcolor}

\usepackage{tikz}
\usetikzlibrary{calc,fadings,decorations.pathreplacing,shapes,shapes.multipart,arrows,shapes.misc,intersections,positioning}

\usepackage{xifthen}
\usepackage{xargs}

\newcommandx{\fineq}[4][1=-.8ex,2=1,3=1]{
  \begin{tikzpicture}[baseline={([yshift=#1]current  bounding  box.center)}, scale = #2, every node/.style={scale = #3}]
    #4
  \end{tikzpicture}
}

\newcommandx{\gate}[3][1=0,2=0,3=]{
  \begin{scope}[shift={(#1,#2)}]
    \ifthenelse{\equal{#3}{}}{
      \draw[line width = 2pt] (0,0)--++(0,2);
      \draw[line width = 2pt] (1,0)--++(0,2);
      \fill[blue!50] (-0.4,0.25) rectangle (1.4,0.85);
      \fill[red!50] (-0.25,1.05) rectangle (0.25,1.25);
      \fill[green!50] (-0.25,1.35) rectangle (0.25,1.55);
      \fill[red!50] (0.75,1.05) rectangle (1.25,1.25);
      \fill[green!50] (0.75,1.35) rectangle (1.25,1.55);
    }{}
    \ifthenelse{\equal{#3}{l}}{
       \draw[line width = 2pt] (0,0)--++(0,2);
       \shade[left color = blue!50, right color = blue!50] (-0.4,0.25) rectangle (.4,0.85);
       \fill[red!50] (-0.25,1.05) rectangle (0.25,1.25);
       \fill[green!50] (-0.25,1.35) rectangle (0.25,1.55);
    }{}
    \ifthenelse{\equal{#3}{r}}{
       \draw[line width = 2pt] (0,0)--++(0,2);
       \shade[left color = blue!50, right color = blue!50] (-0.4,0.25) rectangle (.4,0.85);
       \fill[red!50] (-0.25,1.05) rectangle (0.25,1.25);
       \fill[green!50] (-0.25,1.35) rectangle (0.25,1.55);
    }{}
  \end{scope}
}

\usepackage[english]{babel}
\makeatletter
\def\bbl@set@language#1{%
  \edef\languagename{%
    \ifnum\escapechar=\expandafter`\string#1\@empty
    \else\string#1\@empty\fi}%
  \@ifundefined{babel@language@alias@\languagename}{}{%
    \edef\languagename{\@nameuse{babel@language@alias@\languagename}}%
  }%
  \select@language{\languagename}%
  \expandafter\ifx\csname date\languagename\endcsname\relax\else
    \if@filesw
      \protected@write\@auxout{}{\string\select@language{\languagename}}%
      \bbl@for\bbl@tempa\BabelContentsFiles{%
        \addtocontents{\bbl@tempa}{\xstring\select@language{\languagename}}}%
      \bbl@usehooks{write}{}%
    \fi
  \fi}
\newcommand{\DeclareLanguageAlias}[2]{%
  \global\@namedef{babel@language@alias@#1}{#2}%
}
\makeatother
\DeclareLanguageAlias{en}{english}

\newcommand{\red}{}

\newcommand{\I}{\mathbb{I}}
\usepackage{bm}
\renewcommand{\vec}[1]{\boldsymbol{\mathbf{#1}}}

\newcommand{\tr}[0]{\text{tr}}

\newcommand\tabmultiline[1]{\begin{tabular}{@{}c@{}} #1 \end{tabular}}

\usepackage[colorlinks=true,citecolor=blue,linkcolor=magenta]{hyperref}

\graphicspath{{./}{./images/}}

\begin{document}

\title{Non-unitary Entanglement Dynamics in Continuous Variable Systems}

\author{Tianci Zhou}
\thanks{Present address: Center for Theoretical Physics, Massachusetts Institute of Technology,
Cambridge, Massachusetts 02139, USA}
\email{tzhou13@mit.edu}
\affiliation{Kavli Institute for Theoretical Physics, University of California, Santa Barbara, California 93106, USA}
\author{Xiao Chen}
\email{chenaad@bc.edu }
\affiliation{Department of Physics, Boston College, Chestnut Hill, Massachusetts 02467, USA}
 
\date{\today}

\begin{abstract}
We construct a random unitary Gaussian circuit for continuous-variable (CV) systems subject to Gaussian measurements. We show that when the measurement rate is nonzero, the steady state entanglement entropy saturates to an area-law scaling. This is different from a many-body qubit system, where a generic entanglement transition is widely expected. 
{\red Due to the unbounded local Hilbert space, the time scale to destroy entanglement is always much shorter than the one to build it, while a balance could be achieved for a finite local Hilbert space.}  
By the same reasoning, the absence of transition should also hold for other non-unitary Gaussian CV dynamics.
\end{abstract}

\maketitle

\section{Introduction}
\label{sec:intro}

Recent years have seen a surge of interest in many-body non-unitary quantum dynamics from the perspective of quantum trajectories\cite{cao_entanglement_2018,fan_self-organized_2020,skinner_measurement-induced_2019,bao_theory_2020,jian_criticality_2020,chan_unitary-projective_2019,li_quantum_2018-1,gullans_dynamical_2020-1,li_measurement-driven_2019-1,chen_emergent_2020,li_statistical_2020,lavasani_measurement-induced_2020,sang_measurement_2020,gullans_quantum_2020,Choi_2020}. A simple toy model is a hybrid random circuit composed of both local unitary gates and projective measurements\cite{skinner_measurement-induced_2019,chan_unitary-projective_2019,fan_self-organized_2020,Choi_2020,li_quantum_2018-1,gullans_dynamical_2020-1}. The competition between the unitary dynamics and the local measurement leads to an entanglement phase transition of the steady state, separating the highly entangled volume law phase from the disentangled area law phase\cite{skinner_measurement-induced_2019,chan_unitary-projective_2019,li_quantum_2018-1,gullans_dynamical_2020-1,li_measurement-driven_2019-1}. This discovery leads to a series of developments and discoveries on non-unitary dynamics, such as the error correcting properties of the volume law phase\cite{li_statistical_2020,gullans_quantum_2020,fan_self-organized_2020,Choi_2020}, the symmetry protected non-trivial area law phase\cite{sang_measurement_2020,lavasani_measurement-induced_2020,Ippoliti_2021} and the connection with classical statistical mechanics models\cite{nahum_quantum_2017,zhou_emergent_2019,bao_theory_2020,jian_criticality_2020}.

In all these studies, the local degree of freedom is a qubit or a generalized qudit, which is discrete in nature and these systems are referred to as discrete-variable (DV) systems. In contrast to this, many quantum systems are intrinsically continuous and are referred to as continuous-variable (CV) systems (see review \cite{weedbrook_gaussian_2012}). These systems have infinite Hilbert space dimensions with the physical observables having a continuum of eigenvalues. Furthermore, these CV modes can be coupled together and form a many-body quantum system. In this Research Letter, we build up a Gaussian many-body quantum circuit. In particular, we focus on the non-unitary random circuit models and explore the entanglement scaling of the Gaussian states. {\red Similar to the fermionic Gaussian models explored by others \cite{cao_entanglement_2019,chen_emergent_2020,alberton_entanglement_2021,chen_emergent_2020,jian_criticality_2020,turkeshi_measurement-induced_2021,Biella_2021}, the Gaussian feature brings in tractability, but the physics is different. }

Previously, Ref.~\onlinecite{Zhuang_2019} constructed a Gaussian random unitary circuit by using a set of fundamental one or two-mode CV gates and studied the information spreading and entanglement dynamics of the circuit. They found that analogous to the DV systems, there is a linear light cone for information spreading for systems with local interaction. After the local information reaches the boundary of the system, the entanglement entropy scales linearly in the subsystem size. However, due to the infinite local Hilbert space dimension, the entanglement entropy of a subsystem is unbounded and continues to grow in time. This is different from the aforementioned DV system in which the local degree of freedom takes only $O(1)$ time to reach equilibrium. Such a difference leads to a qualitative change in entanglement dynamics in the hybrid CV non-unitary dynamics. The unitary evolution takes an infinitely long time to entangle the local mode with the rest of the system and cannot compete with the Gaussian measurement which typically disentangles a single mode from the system in $O(1)$ time. As a consequence, there is no entanglement transition in this model and a nonzero measurement rate drives the system to an area law phase. We verify this result numerically in a random Gaussian circuit subject to measurement and generalize this result to other non-unitary CV dynamics (e.g. the CV network system in Ref.~\onlinecite{zhang_entanglement_2021}).


\section{Entanglement dynamics in Gaussian CV systems}
\label{sec:cv_intro}


In this section, we briefly review the quantum information aspect of the continuous-variable systems. We mostly follow the convention of Ref.~\onlinecite{adesso_continuous_2014} with the exception that covariance matrices match $\frac{1}{2}$ of the counterparts in Ref.~\onlinecite{adesso_continuous_2014}. 

A continuous-variable system is a quantum system that has operators with a continuous spectrum. An $L$-mode harmonic oscillator system on a lattice is a prototypical bosonic example. It has position and momentum operators with a continuous spectrum, which are collectively denoted as $X = [\vec{q}, \vec{p}]^{\top}$ (the quadratures). The infinite dimensional Hilbert space can be constructed in the number basis, i.e. the simultaneous eigenstates of the number operators $\prod_{i=1}^N a_i^{\dagger} a_i$, where the mode creation and annihilation operators are related to the quadratures through
\begin{equation}
a_i = \frac{1}{\sqrt{2}} ( q_i + i p_i ) \quad a^{\dagger}_i = \frac{1}{\sqrt{2}} ( q_i - i p_i ) 
\end{equation}
in the convention $\hbar = 1$. These operators have the canonical commutation relations
\begin{equation}
[X_i, X_j]  = i J_{ij}, \quad [a_i, a_j^{\dagger} ] = \delta_{ij} ,
\end{equation}
where $J$ is a $2L \times 2L$ symplectic matrix($\I_L$ is the $L\times L$ identity matrix)
\begin{equation}
J = 
\begin{bmatrix}
0 & \I_{L} \\
-\I_L & 0 \\
\end{bmatrix}.
\end{equation}

Below we review elementary properties of the Gaussian states and Gaussian operations.

\subsection{Gaussian states and Gaussian operations}

{\it Gaussian states} are defined by a Gaussian characteristic or Wigner function. For a pure state, the definition is equivalent to a Gaussian wavefunction in the quadrature (either position or momentum) basis. Its properties are therefore completely determined by the first two moments: the displacement vector $\tr( \rho X )$ and the covariance matrix
\begin{equation}
M_{ij} = \frac{1}{2} \tr( \rho \{ (X_i - \langle X_i \rangle  ), ( X_j - \langle X_j \rangle  ) \} ),
\end{equation}
where $\{,\}$ denotes an anti-commutator. In an $L$-mode system, $M$ is a $2L \times 2L$ real symmetric matrix satisfying the uncertainty principle $2M \ge iJ$\footnote{That is, $2M -i J$ is semi-positive definite.}. It can be ``diagonalized" by a symplectic matrix $S \in \text{Sp}( 2L, \mathbb{R} )$ into the Williamson normal form 
\begin{equation}
  M = S \,\, \text{diag}( \nu_1, \cdots, \nu_L, \nu_1, \cdots, \nu_L ) \,\, S^{\top},
\end{equation}
where $\{ \nu_{i}| i = 1, \cdots, L \}$ are the $L$ (non-negative) Williamson eigenvalues. The R\'enyi entanglement entropies of the Gaussian states only depend on the Williamson eigenvalues and are given by \cite{adesso_continuous_2014}
\begin{equation}
  S_{\alpha} = \frac{\sum_{i = 1}^{L} \ln [ ( \nu_i + \frac{1}{2} )^{\alpha} - ( \nu_i - \frac{1}{2} )^{\alpha } ] }{\alpha - 1}.
\end{equation}
A special example is the second R\'enyi entropy, which is
\begin{equation}
S_2 = \sum_{i=1}^L \ln 2 \nu_i = \frac{1}{2}  \ln \det( 2 M ) . 
\end{equation}
Due to its irrelevance to the entanglement, we set the displacement vectors to be zero afterwards. 

{\it Gaussian operations} transform a Gaussian state to a Gaussian state. For instance, given a Gaussian initial state $|\psi \rangle $, a unitary evolution $e^{- i H t} | \psi \rangle $ with Hamiltonian quadratic in the quadrature (or creation and annihilation operators) produces another Gaussian state and thus a Gaussian operation. Such unitary transformation preserves the commutation relation of the quadratures $[\hat{U}^{\dagger} X_i \hat{U}, \hat{U}^{\dagger} X_j \hat{U}] = [X_i, X_j]$. Infinitesimally, the transformation on the quadrature is linear, which generates a symplectic transformation for finite time: 
\begin{equation}
\hat{U}^{\dagger} X \hat{U} =  S X, \quad S \in \text{Sp}(2L, \mathbb{R}).
\end{equation}
Consequently the Gaussian property of the state is preserved, and the time evolved correlation matrix is given by 
\begin{equation}
M(t) = S M( t = 0 ) S^{\top}. 
\end{equation}

Examples of one-mode and two-mode evolutions are listed in Table~\ref{tab:mode_gate}. We will use these operations as quantum unitary gates in the design of the circuit in Sec.~\ref{sec:mea_circ}.
\begin{table}
\centering
\begin{tabular}{ |c|c|c| } 
 \hline
  Gate & $\hat{u}$ & $S$  \\\hline
  Phase rotation & $\exp( i \theta a^{\dagger} a ) $ & $\begin{bmatrix} \cos \theta & \sin \theta \\  - \sin \theta & \cos \theta \\  \end{bmatrix}$ \\ \hline 
  \tabmultiline{One-mode\\squeezing} & $\exp(  \frac{1}{2} ( r a^\dagger{}^2 - r  a^2 )  )$  & $\begin{bmatrix} e^r  & 0 \\ 0 & e^{-r} \\ \end{bmatrix} $ \\ \hline
  Beam splitter & $\exp(\phi ( a^{\dagger} b - a b^{\dagger} ) )$ & $(\cos \phi \I_2 + i\sin\phi \sigma_y) \otimes \I_2   $\\ \hline 
  \tabmultiline{Two-mode \\ squeezing} & $\exp(r(  a^{\dagger} b^{\dagger}  - a b ) )$& $ \cosh r \I_4 + \sinh \sigma_z \otimes \sigma_x$\\ \hline
\end{tabular}
\caption{Unitary gates on one mode and two modes. $\hat{u}$ is the unitary operator in terms of the boson creation and annihilation operators. $S$ is the corresponding symplectic matrix, which acts on the quadrature, $[q_1, p_1]^{\top}$ for one mode and $[q_1, p_1, q_2, p_2]^{\top}$ for two modes. }
\label{tab:mode_gate}
\end{table}

Furthermore, certain measurements are Gaussian operations. For example, a homodyne measurement corresponds to a measurement that projects to quadrature basis $| q \rangle $ or $| p \rangle $ (infinitely squeezed states). It can produce a Gaussian state for the part of the system that is not measured. Projection to a coherent state $| \alpha \rangle  = D( \alpha ) |0 \rangle $ (a displaced harmonic oscillator ground state, $D(\alpha) = \exp( \alpha a^\dagger - \bar{\alpha} a )$ is the displacement operator) is called a heterodyne measurement. For pure states, since the wavefunctions are Gaussian, their overlaps with another Gaussian wavefunction are again Gaussian. Hence projection to Gaussian states are Gaussian operations. A similar argument can be made for a mixed state, in which we can first purify to a larger Gaussian state, then take the measurement, and finally trace out the environment. 

Finally, imaginary time evolution as {\red a weak measurement with post-selection} is also a Gaussian operation. This means that the normalized state under an imaginary time evolution of a quadratic Hamiltonian $H$
\begin{equation}
\frac{e^{ - \beta H}  | \psi \rangle }{\langle \psi |e^{-2 \beta H} |\psi  \rangle }
\end{equation}
is a Gaussian state if the original state $|\psi \rangle $ is (for a proof see \cite{SM}).
This is in parallel with several earlier works\cite{cao_entanglement_2019,chen_emergent_2020,alberton_entanglement_2021,jian_criticality_2020} on the Gaussian formalism for the measurement-based transition in fermionic systems.

In Sec.~\ref{sec:mea_circ}, we will construct an $L$-mode hybrid quantum circuit model by applying a sequence of the one-mode and two-mode Gaussian gates introduced in this section. 

\subsection{Entanglement for Two-mode System}
\label{subsec:ee-two-mode}
Qubit systems have a finite local Hilbert space, and entanglement of a finite spatial region has a saturation value upper bounded by the logarithm of the Hilbert space dimension. This is not the case for a CV system. The infinite local Hilbert space gives rise to an unbounded entanglement growth, even for a two-mode system. 

For example, starting from the vacuum state we repeatedly apply a two-mode squeezing gate with squeezing parameter $r$ to obtain the state,
\begin{equation}
\begin{aligned}
 | \psi(t) \rangle  &= [\exp( r( a^{\dagger} b^{\dagger} - ab ) )]^t |0 0 \rangle  \\
 & = \frac{1}{ \cosh rt } \sum_{n=0}^{\infty} (\tanh rt )^n  | n n \rangle.
\end{aligned}
\end{equation}
The second R\'enyi entropy for the subsystem of the first mode can be calculated from the covariance matrix $M = \frac{1}{2} S S^{\top} $. We have 
\begin{equation}
S_2 =  \ln \cosh 2 rt  \sim 2rt  \text{ for large } rt.
\end{equation}
The entanglement has an unbounded growth with an asymptotic linear time dependence.

This state has the property
\begin{equation}
\langle  (q_1 - q_2)^2 \rangle  = \langle  ( p_1 + p_2 )^2 \rangle  = e^{-rt }.
\end{equation}
When $rt$ is large, it is a very good practical approximation of the perfect Einstein-Podolsky-Rosen(EPR) state with two photons at the same position and having almost opposite momentum
\begin{equation}
 \delta ( q_1 - q_2 ) \delta ( p_1 + p_2 ). 
\end{equation}


\section{Hybrid Gaussian CV Circuit}
\label{sec:mea_circ}


In a qubit system, it is expected that there is a generic entanglement phase transition in the hybrid quantum circuit composed of both unitary evolution and projective measurements\cite{skinner_measurement-induced_2019,chan_unitary-projective_2019,fan_self-organized_2020,Choi_2020,li_quantum_2018-1,gullans_dynamical_2020-1}. When the measurement rate is small, there is a stable volume law phase. As we increase the measurement rate, there is a transition to the disentangled area law phase.

\begin{figure}[h]
\centering
\includegraphics[width=0.9\columnwidth]{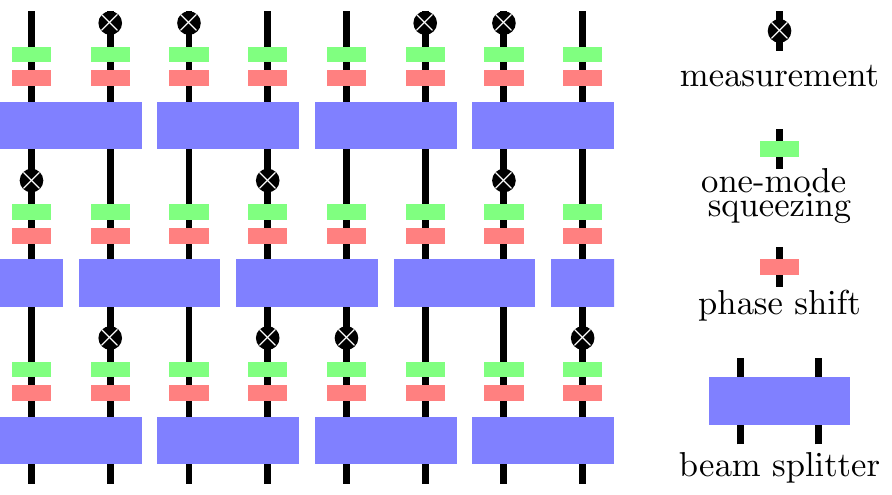}
\caption{Structure of the hybrid circuit. Random one-mode squeezing and phase shift are used to scramble the local Hilbert space, and a random two-mode beam splitter is used to entangle nearby modes. Measurements to the vacuum state are inserted with rate $p$. }
\label{fig:mea_circ}
\end{figure}

Inspired by the developments in the DV system, in this section, we construct a similar circuit for a CV system; see Fig.~\ref{fig:mea_circ}.
We use independent random phase shift and one-mode squeezing to scramble the local Hilbert space and a random beam splitter to entangle neighboring sites. The random parameter choices are uniform random $\theta$ and $\phi$ in $[0, 2\pi]$ and uniform random squeezing parameter $r \in [0,1]$, see explicit expressions in Tab.~\ref{tab:mode_gate}. Then with probability $p$, {\red we apply the measurement and post-select the outcome to be the single mode ground state $|0 \rangle $ at each site. This is a measurement with a forced outcome.}
The initial state is chosen as the tensor products of $|0 \rangle $. 

\begin{figure*}[t]
\centering
\subfigure[]{
  \label{fig:S2_t2}	
  \includegraphics[width=0.31\textwidth]{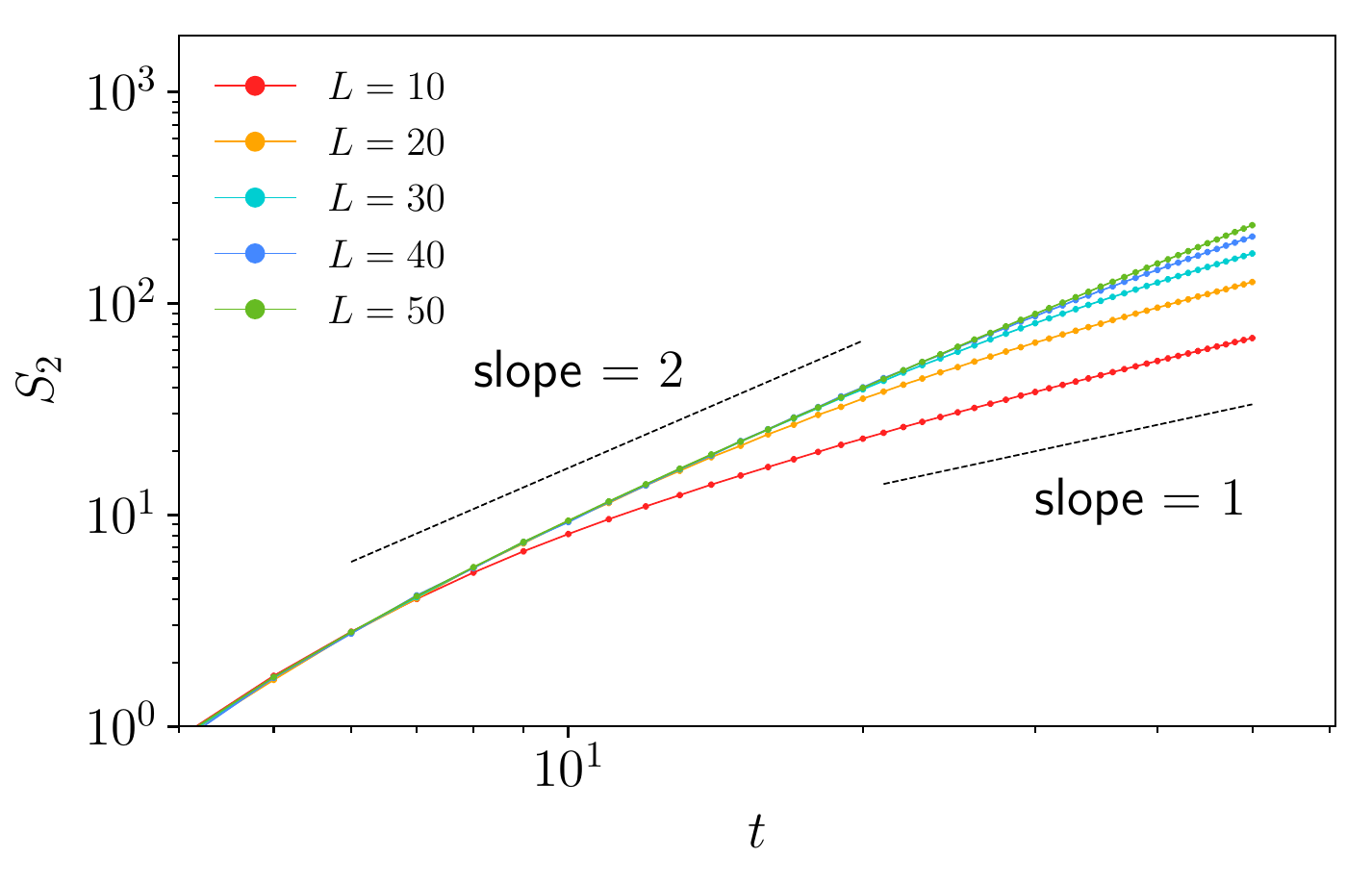}
}
\subfigure[]{
  \label{fig:hetero_pL}	
  \includegraphics[width=0.31\textwidth]{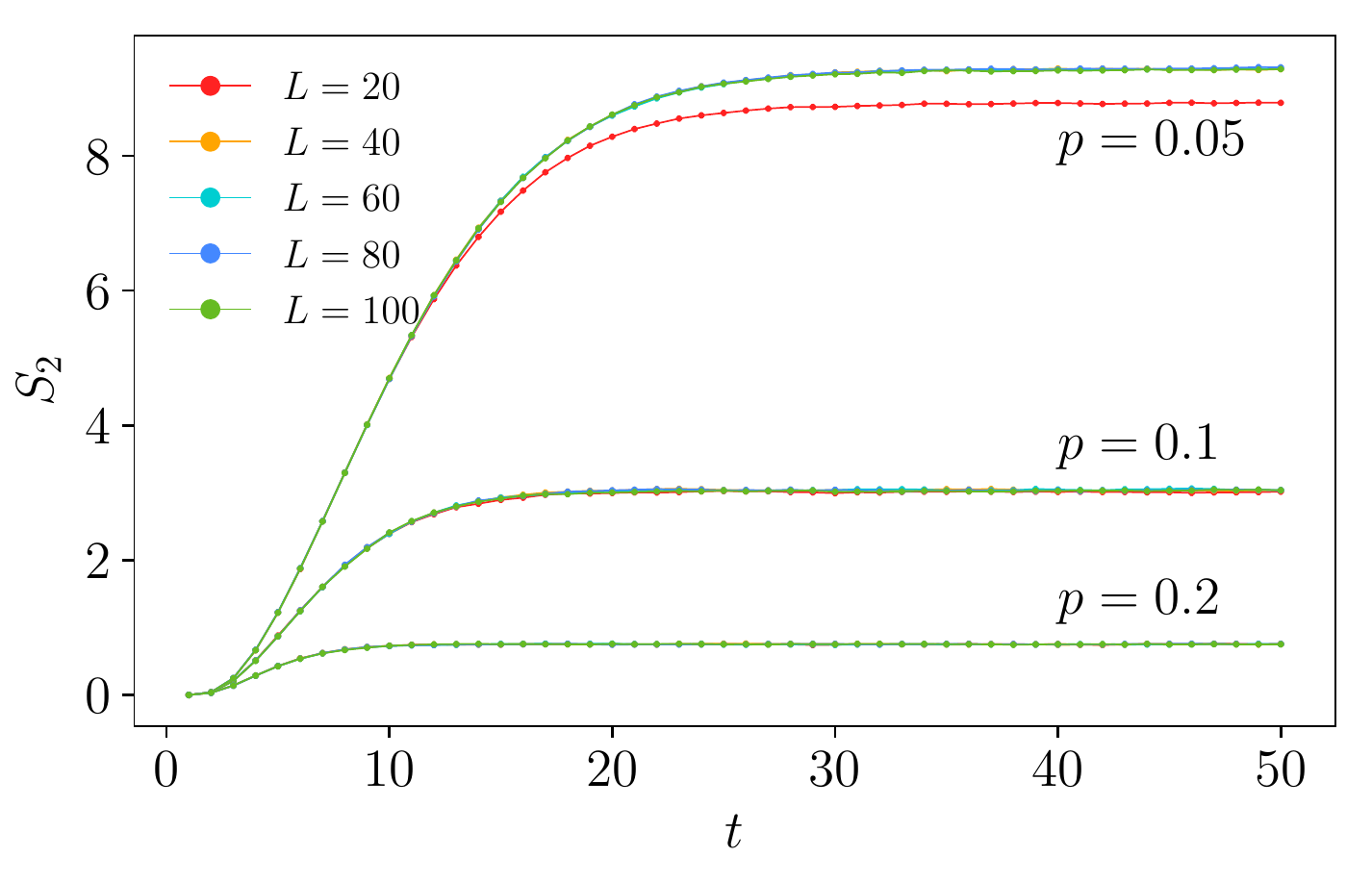}
}
\subfigure[]{
  \label{fig:hetero_betaL}	
  \includegraphics[width=0.31\textwidth]{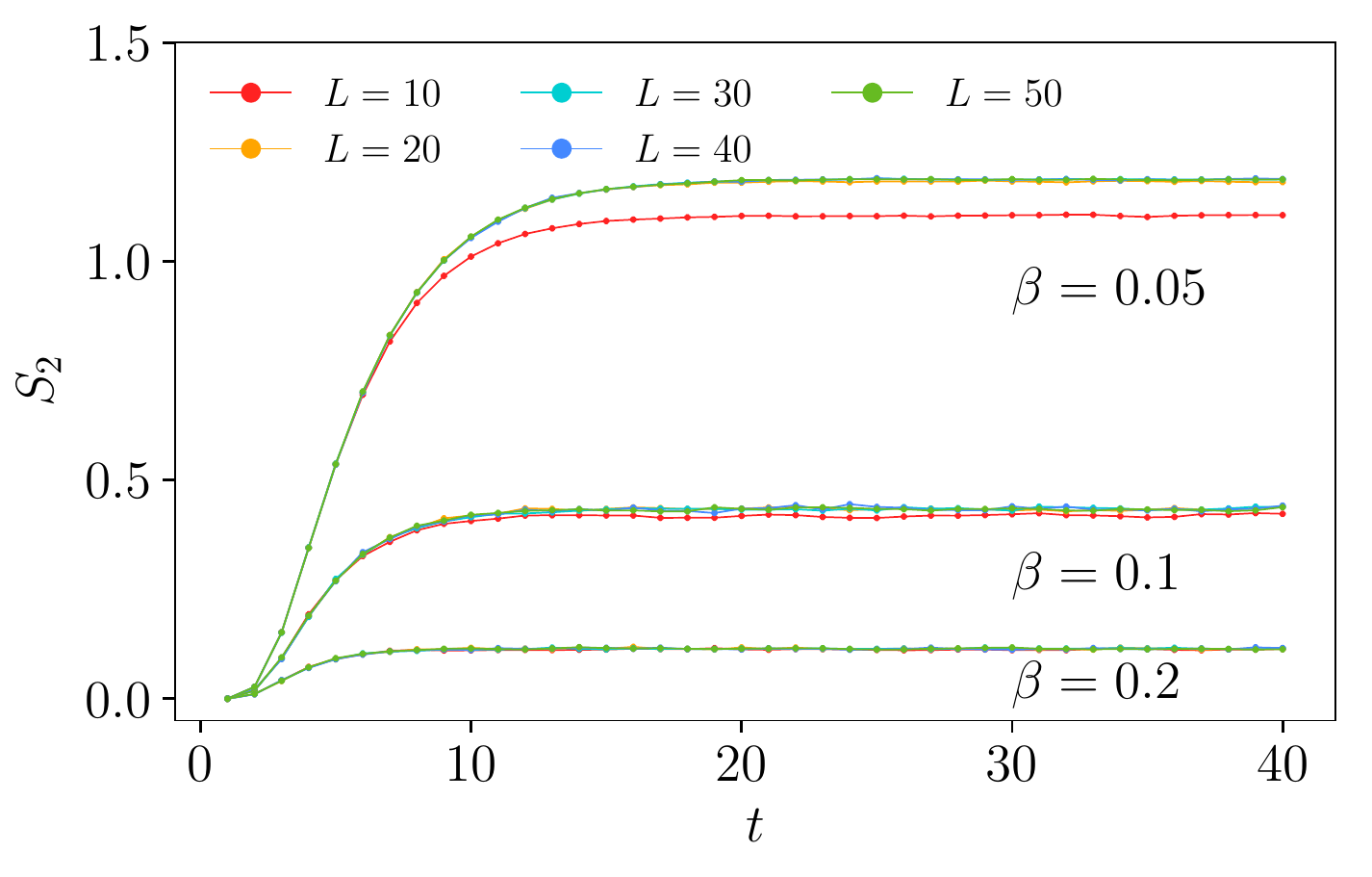}
}
\caption{Averaged half system entanglement $S_2$ in a hybrid circuit. Entanglement is averaged over random realizations of the gate. There are two circuit structures: applying gates either on odd bonds in the first step or on even bonds. The results here are the averages over equal probabilities of these two structures.  (a) When the measurement probability $p = 0$, entanglement has a quadratic growth followed by a linear growth. (b) When $p > 0$, the averaged entanglement converges to system size independent values, with a slightly smaller saturation value when $L \lesssim \frac{1}{p} $. (c) We replace the Gaussian measurement by the imaginary evolution gate $e^{- \beta \sum_i a_i^\dagger a_i } $. We set $p=1$ and vary the strength parameter $\beta$. The entanglement converges to system size independent values. }
\label{fig:mea_res}
\end{figure*}
\twocolumngrid

When $p = 0$, the circuit is completely unitary; such a circuit has been studied in Ref.~\onlinecite{Zhuang_2019}. In a DV system with local interactions, the von Neumann entanglement entropy grows linearly in time and saturates to a volume law scaling\cite{Kim_2013}. The situation is quite different here in the CV setting. The entanglement entropy has an initial $t^2$ growth, and then crosses over to a linear unbounded growth, see Fig.~\ref{fig:S2_t2}.

The entanglement can be estimated by computing the dimension of the effective Hilbert space. In a DV system, the maximal local Hilbert space dimension is fixed to be $d$. The locally interacting gate expands the domain of influence linearly. Hence the effective Hilbert space dimension along either side of the entanglement cut is roughly $d^t$. Taking the logarithm gives a linear growth of the entropy. After the saturation time, $t \sim L$, the entanglement saturates to a volume law value. In a CV system, besides the linear spreading in the spatial direction, the explored local Hilbert space dimension $d$ is also growing exponentially in time {\red (see the linear growth of entanglement worked out in Sec.~\ref{subsec:ee-two-mode})}. Taking into account both effects gives a $t^2$ growth at early time (see Fig.~\ref{fig:S2_t2}). After the domain of influence reaches the whole system, the effective Hilbert space can continue to grow as $d^L \sim e^{t L}$. Taking the logarithm gives rise to the late time linear $t$ growth observed in Fig.~\ref{fig:S2_t2}.

When we turn on measurements that project to $|0 \rangle$ with finite rate of measurement $p$,
the entanglement growth is system size independent (see \cite{SM} for the numerical algorithm). It has a short time growth and eventually saturates to a system size independent value, i.e. an area law, see Fig.~\ref{fig:hetero_pL}.

We replace the measurement by an imaginary single-mode gate. {\red This is a continuous weak measurement with post-selection.} Operationally, we set $p = 1$ (measuring all sites after the one-mode squeezing) and replace the measurement from the projection to $|0 \rangle $ to a weaker imaginary time evolution $e^{ - \beta a^{\dagger}_i a_i }$. The tuning parameter is now $\beta$, {\red which indicates the strength of the measurement.} We observe similar area law behavior in Fig.~\ref{fig:hetero_betaL} when $\beta$ is finite.  

We believe that the infinite Hilbert space dimension plays the crucial role in the absence of the entanglement phase transition. For the sake of argument, assume that the local Hilbert space dimension for the CV system is truncated at a finite but very large number $K$. It takes $O(\ln K)$ time to explore the local Hilbert space and creates entanglement $\ln K$, whereas a single Gaussian measurement can destroy this entanglement immediately.  As $K \rightarrow \infty$, the time scale associated with the measurement that destroys the entanglement is much smaller than the time scale to create entanglement, hence pushing the would-be transition probability $p_c$ to $0$.

\subsection{DV dynamics example}

\begin{figure}[t]
 \includegraphics[width=1\columnwidth]{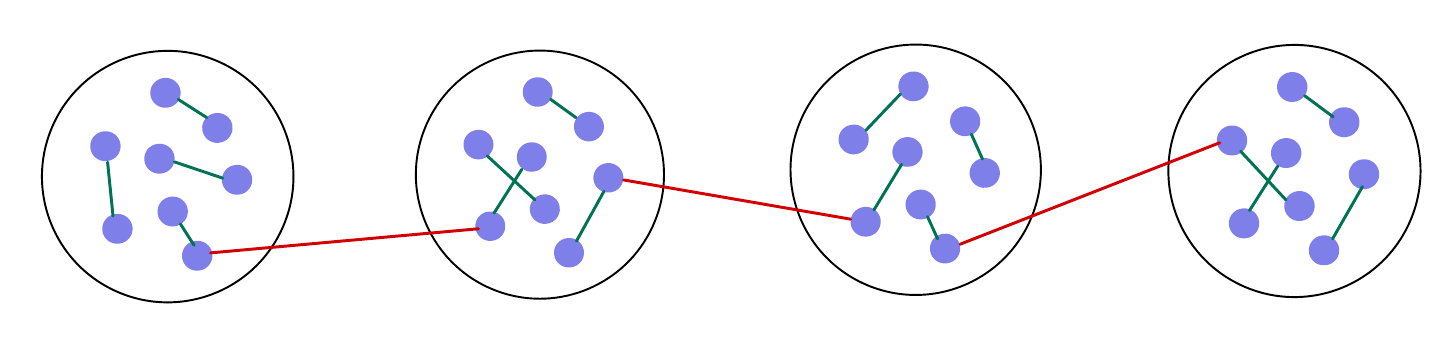}
\caption{A cartoon of the qubit system. At each site, there are $N$ qubits. The unitary dynamics at each time step involves two parts: (1) intra-site interaction described by the $N/2$ two-qubit gates (denoted by the green lines), and (2) a single two-qubit gate connecting neighboring sites (red lines). } 
\label{fig:DV}
\end{figure}

We can reproduce similar physics in a DV model with large local Hilbert space dimension. Consider a one dimensional qubit system with $L$ sites. At each site, there is a cluster of $N$ qubits. As shown in Fig.~\ref{fig:DV}, in each time step, the unitary evolution involves both intra-cluster interaction and inter-cluster interaction. The intra-cluster interaction is realized by applying $N/2$ two-qubit gates which randomly couple $N/2$ pairs of qubits, while the inter-cluster interaction between two neighboring sites only has a {\it single} two-qubit gate. The projective measurement gate is applied at each site with probability $p$ and can disentangle {\it every qubit} in one cluster. 

The design of the interaction patterns ensure that there is roughly a $\ln 2$ entanglement increase across the inter-cluster bond, until it reaches the maximum of $N \ln 2$. Hence it takes $O(N)$ time for one cluster to get fully entangled with other clusters under unitary evolution. However, it takes only $O(1)$ time for one cluster to become disentangled under the projective measurement\footnote{In a Haar random circuit, the absence of the phase transition in the large $N$ limit can also be understood in terms of the minimal cut picture\cite{skinner_measurement-induced_2019}.}. This is comparable to the $\ln K$ time scale (note that $K$ is the local Hilbert space dimension in the CV system) to entangle and $O(1)$ time to disentangle in the CV case.

At finite $N$, the two time scales are still comparable, resulting in a phase transition at finite $p_c$. However, $p_c$ will decrease with $N$ and eventually vanishes when $N \rightarrow \infty$. 

\begin{figure}[t]
 \includegraphics[width=0.9\columnwidth]{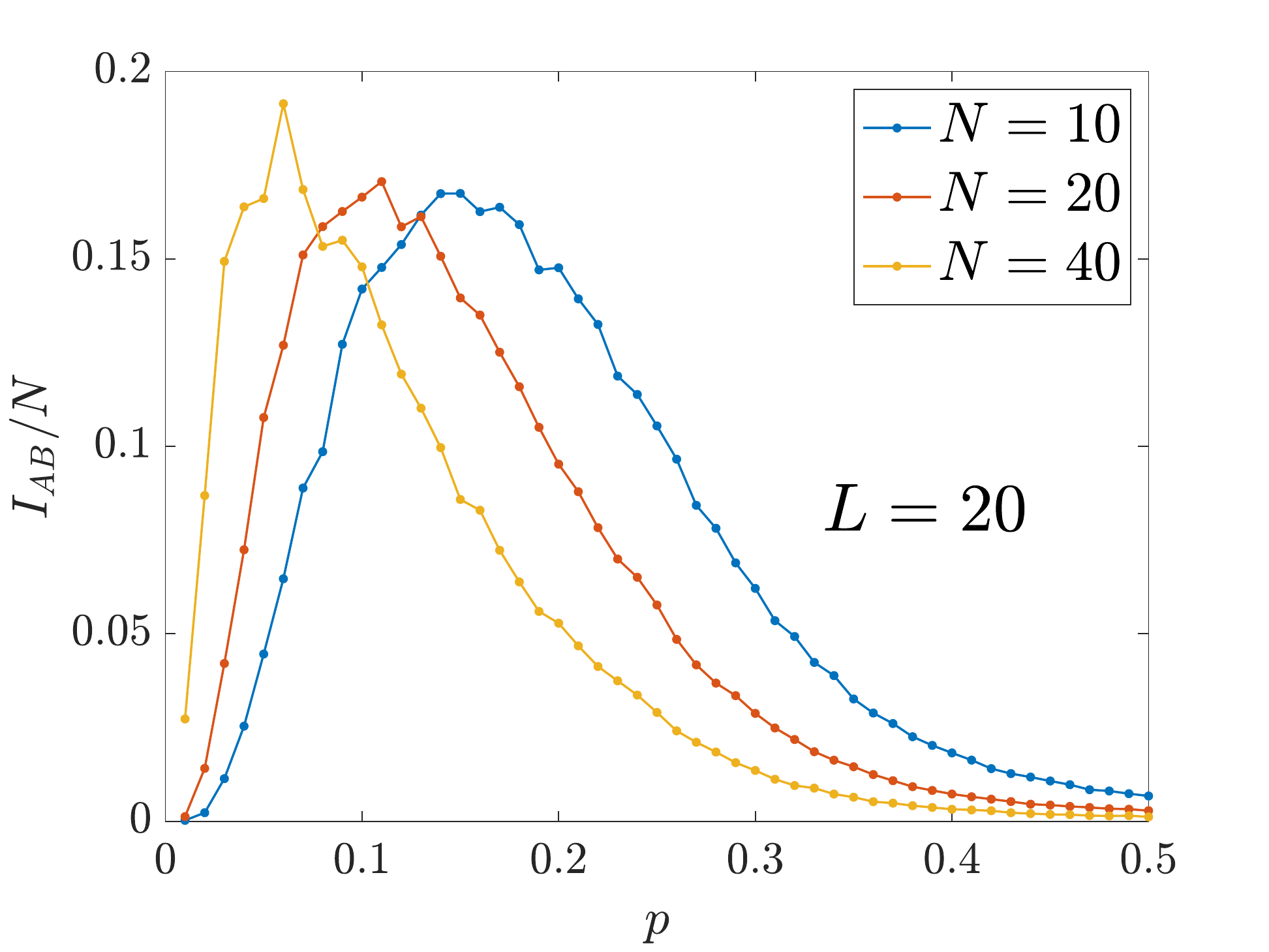}
\caption{We consider a hybrid Clifford circuit composed of both unitary dynamics and projective measurement and compute the steady state mutual information $I_{AB}=S_{A}+S_{B}-S_{AB}$ for various $N$. At each time step, the unitary dynamics is described by Fig.~\ref{fig:DV} with each bond denoting a random Clifford gate, and the projective measurement is applied randomly at each site with the probability $p$. After the measurement, all the qubits at one site are disentangled from the rest of system. At finite $N$, there is an entanglement phase transition with $p_c$ characterized by the peak of $I_{AB}$ of the steady state\cite{li_measurement-driven_2019-1}. Here we consider periodic boundary conditions and $A$ and $B$ are two antipodal regions with length $L_A=L_B=4$.}
\label{fig:Clifford}
\end{figure}

We numerically verify this idea in a random Clifford circuit and present the result in Fig.~\ref{fig:Clifford}. This model can be efficiently simulated with a large number of qubits\cite{gottesman1998heisenberg,Aaronson_2004}. We use the peak of the mutual information to identify the location of $p_c$\cite{li_measurement-driven_2019-1}. As $N$ increases, $p_c$ moves to the left and approaches zero in the large $N$ limit. 

Notice that if we modify the unitary dynamics in Fig.~\ref{fig:DV} and introduce $N$ two-qubit gates connecting neighboring sites, then the time scale to completely entangle two clusters reduces to $O(1)$ and there is an entanglement phase transition at finite $p_c$ \cite{liu2021nonunitary}, even after sending $N\rightarrow \infty$.

\section{Conclusion}
\label{sec:conclusion}


In this Research Letter, we study the entanglement dynamics in hybrid CV Gaussian circuits composed of both unitary and non-unitary gates. For a generic random unitary Gaussian dynamics, the entanglement entropy for a subsystem is proportional to the subsystem size and can grow indefinitely in time due to the unbounded local Hilbert space dimension. We show that this highly entangled phase is unstable when the system is subject to repeated Gaussian measurements. When the measurement rate is nonzero, the steady state evolves to an area law phase, indicating the absence of the entanglement transition.

We argue that the lack of phase transition is due to disparity of the competing time scales to entangle and disentangle the degrees of freedom. While it takes an $O(1)$ time scale to destroy the entanglement in a Gaussian measurement, it takes an infinitely long time for a single mode to get entangled with the system. We reproduce this effect from a similar DV system construction with finite but large local Hilbert space dimension.

Our result holds for a generic Gaussian unitary circuit subject to Gaussian measurements and can be applied to other hybrid Gaussian dynamics, in which the Gaussian measurement is replaced by an imaginary evolution gate. We expect that our argument for the absence of the entanglement transition can be generalized to interacting hybrid CV dynamics. Moreover, it might be interesting to explore hybrid CV dynamics with extra constraints, such as global ${\rm U}(1)$ symmetry, where an entanglement phase transition may exist. We leave this for future study.


\acknowledgements

T.Z. was supported by a postdoctoral fellowship from the Gordon and Betty Moore Foundation, under the EPiQS initiative, Grant No. GBMF4304, at the Kavli Institute for Theoretical Physics. 
This research is supported in part by the National Science Foundation under Grant No. NSF PHY-1748958.
We acknowledge support from the Center for Scientific Computing from the CNSI, MRL--an NSF MRSEC (Award No. DMR-1720256) and from NSF Award No. CNS-1725797.

\bibliographystyle{apsrev4-1}
\bibliography{CV_mea_paper}

\end{document}


\title{Supplemental Material for ``Non-unitary Entanglement Dynamics in Continuous Variable Systems''}

\author{Tianci Zhou}
\thanks{Present address: Center for Theoretical Physics, Massachusetts Institute of Technology,
Cambridge, Massachusetts 02139, USA}
\email{tzhou13@mit.edu}
\affiliation{Kavli Institute for Theoretical Physics, University of California, Santa Barbara, California 93106, USA}
\author{Xiao Chen}
\email{chenaad@bc.edu }
\affiliation{Department of Physics, Boston College, Chestnut Hill, Massachusetts 02467, USA}
 
\date{\today}
\maketitle

\section{Quadratic Imaginary Time Evolution is Gaussian}
\label{app:beta_gau}
\input{app_beta_gau}

\section{Numerical Algorithms}
\label{app:num_algor}
\input{app_num_algor}

\bibliographystyle{apsrev4-1}
\bibliography{CV_mea_paper}